\documentstyle{mn}
\input {psfig.sty}
\def \aj {AJ}
\def \mnras {MNRAS}
\def \apj {ApJ}
\def \apjl {ApJL}
\def \aap {A\&A}
\def \Mpc {~h^{-1}~{\rm Mpc} }
\def \Om {\Omega_0}

\def \lo {\lambda_0}

\def \bj {b_{\rm J}}
\def \qso {_{\rm Q}}
\def \gal {_{\rm g}}
\def \xibar {\bar{\xi}}
\def \max {_{\rm max}}
\def \gsim { \lower .75ex \hbox{$\sim$} \llap{\raise .27ex \hbox{$>$}} }
\def \lsim { \lower .75ex \hbox{$\sim$} \llap{\raise .27ex \hbox{$<$}} }

\title[The 2QZ: luminosity dependence of QSO clustering]
      {The 2dF QSO Redshift Survey - IX.  A measurement of the luminosity dependence of QSO clustering}
\author[S.M. Croom et al.]
       {Scott M. Croom$^{1}$\thanks{scroom@aaoepp.aao.gov.au},
      B.J. Boyle$^1$,  N.S. Loaring$^2$, L. Miller$^2$,
      P.J. Outram$^3$, T. Shanks$^3$ \newauthor R.J. Smith$^4$\\
$^1$Anglo-Australian Observatory, PO Box 296, Epping, NSW 2121,
      Australia.\\
$^2$Department of Physics, Oxford University, Keble Road, Oxford, OX1
      3RH, UK.\\
$^3$Physics Department, University of Durham, South Road, Durham, DH1 3LE,
UK.\\
$^4$Astrophysics Research Institute, Liverpool John Moores University,
Twelve Quays House, Egerton Wharf, Birkenhead,  CH41 1LD, UK.}

\begin{document}

\maketitle

\begin{abstract}
In this Paper we present a clustering analysis of QSOs as a function
of luminosity over the redshift range $z=0.3-2.9$.  We use a sample of
10566 QSOs taken from the preliminary data release catalogue of the
2dF QSO Redshift Survey (2QZ).  We analyse QSO clustering as a
function of {\it apparent} magnitude.  The strong luminosity evolution
of QSOs means that this is approximately equivalent to analysing the
data as a function of absolute magnitude relative to $M^*$ over the
redshift range that the 2QZ probes.  Over the relatively narrow range
in apparent magnitude of the 2QZ we find no significant ($>2\sigma$)
variation in the strength of clustering, however, there is marginal
evidence for QSOs with brighter apparent magnitudes having a stronger
clustering amplitude.  QSOs with $18.25<\bj\leq19.80$ show a
correlation scale length $s_0=5.50\pm0.79\Mpc$ in an Einstein-de
Sitter (EdS) universe and $s_0=8.37\pm1.17\Mpc$ in a universe with
$\Om=0.3$ and $\lo=0.7$ ($\Lambda$), while the best fit values for the
full magnitude interval ($18.25<\bj\leq20.85$) over the same spatial
scales are $s_0=4.29^{+0.30}_{-0.29}\Mpc$ (EdS) and
$s_0=6.35^{+0.45}_{-0.44}\Mpc$ ($\Lambda$).  We can therefore
determine that the bias of the brightest sub-sample is a factor
$1.22\pm0.15$ (EdS) or $1.24\pm0.15$ ($\Lambda$) larger than that of
the full data set.  An increase in clustering with luminosity, if
confirmed, would be in qualitative agreement with models in which the
luminosity of a QSO is correlated to the mass of the dark halo in
which it resides, implying that the mass of the host plays at least
some part in determining a QSO's formation and evolution.  These
models predict that the clustering in brighter QSO data sets, such as
Sloan Digital Sky Survey QSO sample or the bright extension of the 2QZ
should show a higher clustering amplitude than the 2QZ. 
\end{abstract}


\section{Introduction}

The currently preferred models of structure formation based on
hierarchical growth of structure predict that the level of clustering
of a population is dependent on the mass of the dark halo in which the
object resides.  Thus clusters of galaxies should cluster much more
strongly that galaxies do, which we observe to be the case.  Analysis
of galaxy surveys have also shown that more luminous (and therefore on
average more massive) galaxies have stronger clustering than faint
galaxies \cite{gd01,lmep95}.

The luminosity of QSOs appears to be (weakly) correlated with host
galaxy luminosity, at least for low redshift QSOs (e.g. Schade, Boyle
\& Letawsky 2000).  More luminous QSOs reside in brighter host
galaxies.  It is then natural to suppose that brighter QSOs should be
found in more massive galaxies, assuming that luminosity and mass are
correlated at least to some extent.  Further circumstantial support
for this argument is given by the correlation between the estimates of
central black hole mass and the spheroidal component of the galaxies
\cite{mag98}.  We would then expect more luminous QSOs to cluster more
strongly than faint QSOs.

\begin{figure}
\centering
\centerline{\psfig{file=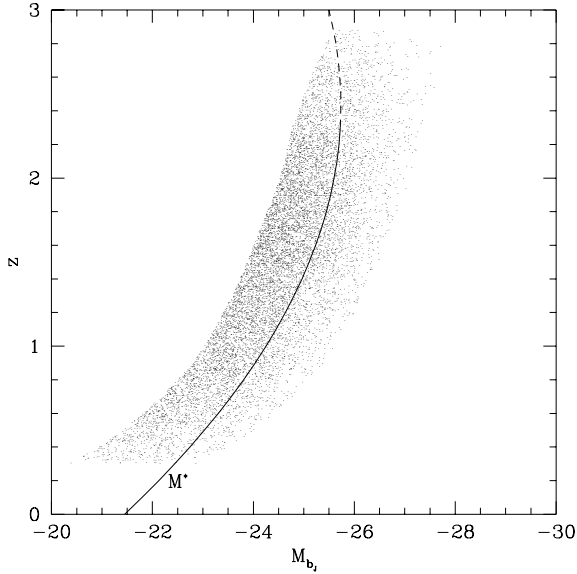,width=8.0cm}}
\caption{The redshift vs. absolute magnitude distribution of 2QZ QSOs
used in our analysis for an EdS cosmology.  The
solid line shows the value of $M^*$ as a function of redshift, as
derived from the best fit LF model (see text).  the model fit is
calculated between $z=0.3$ and 2.3, beyond $z=2.3$ the model is an
extrapolation (dashed line).}
\label{fig_magz}
\end{figure}

Until recently it was impossible to determine QSO clustering as a
function of luminosity.  The sparse nature of QSO surveys and the
small numbers in homogeneous, complete samples meant that QSO
clustering was only detectable at the $\sim4\sigma$ level
\cite{is88,ac92,sb94,cs96,lac98}.  However, the 2dF QSO Redshift
Survey (2QZ) has allowed a dramatic improvement to be made in the
measurement of QSO clustering.  Using over 10000 QSOs from the 2QZ,
Croom et al. 2001a (henceforth Paper II) have made the first accurate
(to $\sim10$ per cent) measurement of QSO clustering.  They find that
QSO clustering averaged over the redshift range $0.3\le z\le2.9$ is
very similar to that measured in redshift surveys of local ($z\sim
0.05$) galaxies.  When fitting a standard power law of the form
$\xi\qso(s)=(s/s_0)^{-\gamma}$ Croom et al. find
$s_0=3.99^{+0.28}_{-0.34}\Mpc$ and $\gamma=1.58^{+0.10}_{-0.09}$ for
an Einstein-de Sitter Universe (henceforth denoted by EdS).  For a
cosmology with $\Om=0.3$ and $\lo=0.7$ (henceforth denoted by
$\Lambda$) they find $s_0=5.69^{+0.42}_{-0.50}\Mpc$ and
$\gamma=1.56^{+0.10}_{-0.09}$.  When investigated as a function of
redshift the clustering of QSOs was found to be constant over the
whole redshift range considered.

The nature of any flux limited sample means that the most distant
objects in the data set will on average have the highest intrinsic
luminosity.  There is then the reasonable concern that a
comparison of QSO or galaxy clustering at different redshifts in any
flux limited sample is actually measuring the properties of a
different population of objects at each redshift.  In the case of
QSOs, their extreme evolution in luminosity [$\propto(1+z)^3$] out to
$z\sim2$ (Boyle et al. 2000; henceforth Paper I) means that at each
redshift we are at least studying the same part of the QSO luminosity
function (LF).

In this paper we will make the first attempt to determine the strength
of QSO clustering as a function of luminosity.  In Section
\ref{sec_data} we describe the data and analysis used.  In Section
\ref{sec_results} we present our clustering results from the
2QZ, these are discussed in Section \ref{sec_discuss}.

\section{Data and analysis}\label{sec_data}

\begin{table*}
\baselineskip=20pt
\begin{center}
\caption{2QZ clustering results as a function of limiting apparent
magnitude.  The fits assume a fixed power law slope equal to that of
the best fit from the full data set.  The listed $\chi^2$ value is a
reduced $\chi^2$ for the best fit.}
\begin{tabular}{ccccccccc}
\hline
($\Om$,$\lo$) & $\bj$ range & $\bar{\bj}$ & $\bar{z}$ & $\bar{M}_{\rm J}$ & $N\qso$ &  $s_0$ & $\chi^2$ & $\xibar(20)$\\
\hline
(1.0,0.0) & $18.25<\bj\leq19.80$ & 19.29 & 1.398 & --25.05 & 3518 & $5.50^{+0.79}_{-0.79}$ &  1.70 & $0.236\pm0.077$\\
(1.0,0.0) & $19.80<\bj\leq20.40$ & 20.12 & 1.509 & --24.43 & 3697 & $2.83^{+1.04}_{-1.24}$ &  0.93 & $0.171\pm0.072$\\
(1.0,0.0) & $20.40<\bj\leq20.85$ & 20.63 & 1.552 & --23.98 & 3351 & $4.01^{+0.95}_{-0.98}$ &  0.34 & $0.206\pm0.082$\\
(0.3,0.7) & $18.25<\bj\leq19.80$ & 19.29 & 1.398 & --25.70 & 3518 & $8.37^{+1.17}_{-1.17}$ &  0.75 & $0.722\pm0.152$\\
(0.3,0.7) & $19.80<\bj\leq20.40$ & 20.12 & 1.509 & --25.11 & 3697 & $4.05^{+1.59}_{-1.90}$ &  0.76 & $0.317\pm0.132$\\
(0.3,0.7) & $20.40<\bj\leq20.85$ & 20.63 & 1.552 & --24.67 & 3351 & $5.92^{+1.45}_{-1.49}$ &  0.83 & $0.358\pm0.150$\\
\hline
\label{table_fits}
\end{tabular}
\end{center}
\end{table*}

\subsection{The QSO sample}

For the analysis in this paper we have used the first public release
catalogue of the 2QZ, the 10k catalogue (Croom et al. 2001b;
henceforth Paper V).  We only use QSOs with high quality (quality
class 1; see Paper V) identifications of which there 10689 in the 10k
catalogue (note that the number of QSOs used in the analysis of Paper
II was 10681, 8 further QSOs were added to the 10k sample prior to its
final publication).  This 10k catalogue contains the most
spectroscopically complete fields observed prior to November 2000 and
is now publically available to the astronomical community at {\tt
http://www.2dfquasar.org}.  The sample contains 10566 QSOs in the
redshift range $0.3<z\leq2.9$ which will be included in our analysis
below.

The magnitude range of the 2QZ is $18.25<\bj\leq20.85$.  This is a
fairly narrow range, covering just over an order of magnitude in flux.
However it does span the important regime in which the QSO luminosity
function flattens towards fainter magnitudes.  The distribution of
QSO absolute magnitudes verses redshift is shown in
Fig. \ref{fig_magz} for the EdS cosmology.  We convert from
apparent to absolute luminosity using the k-correction of Cristiani \&
Vio (1990) and assume a Hubble constant of
$H_0=50$~km~s$^{-1}$~Mpc$^{-1}$ (for luminosity determinations only).
We note that in our clustering analysis we include the factor $h$
where $H_0=100h^{-1}$~km~s$^{-1}$~Mpc$^{-1}$.  The
objects at high redshift are on average much more luminous, as we
would expect in a flux limited sample.  Also, as the number of QSOs
drops dramatically at brighter apparent luminosities there is only a
small amount of overlap in intrinsic luminosity between QSO samples
selected at widely differing redshifts.  

The QSO LF is well described by a double power
law.  The characteristic absolute magnitude of the break is denoted by
$M^*$ which is found to evolve strongly with redshift.  Using the
power law evolution model of Paper I, in which $M^*(z)=M^*(0)-2.5(k_1
z + k_2 z^2)$, Paper V derived the best fit parameters for the 2QZ 10k
catalogue and Large Bright Quasar Survey \cite{lbqs6} combined in an
EdS universe.  The parameters found were:
$\Phi^*=0.2\times10^{-5}\,$Mpc$^{-3}\,$mag$^{-1}$, $\alpha=3.28$,
$\beta=1.08$, $M_{\rm B}^*=-21.45$, $k_1=1.41$ and $k_2=-0.29$.  $M^*$
as a function of redshift for this model is plotted in
Fig. \ref{fig_magz}.  The fit to the LF is only carried out below
$z=2.3$, above that redshift the dashed line shows the extrapolation
of the best fit model, which is approximately consistent with the
density of high redshift ($3.6\leq z\leq5$) QSOs found in the Sloan
Digital Sky Survey \cite{sdsslf01}.  We have also fit the same model
to the 2QZ and LBQS assuming a $\Lambda$ cosmology.  In this case we
find: $\Phi^*=0.46\times10^{-6}\,$Mpc$^{-3}\,$mag$^{-1}$,
$\alpha=3.42$, $\beta=1.35$, $M_{\rm b}^*=-22.54$, $k_1=1.34$ and
$k_2=-0.26$.  This fit is in the range of acceptability
specified in Paper I ($P_{\rm KS}>0.01$) with a 2-D Kolmogorov-Smirnov
(KS) probability of 0.02.  It is worth noting that the
parameters of these fits are strongly correlated, hence the use of any
single parameter is to be discouraged.  As this is the case we make
comparisons between QSO samples of different {\it apparent}
magnitudes, and do not split the QSOs in luminosity relative to
$M^*$.

\begin{figure}
\centering
\centerline{\psfig{file=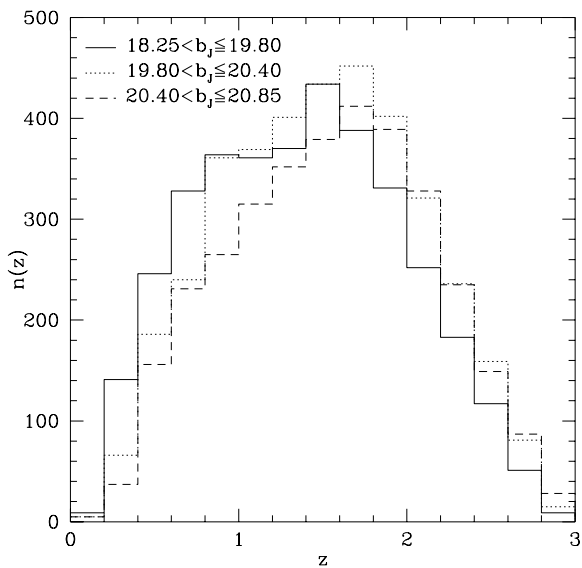,width=8.0cm}}
\caption{The QSO redshift distribution for our three apparent
magnitude intervals, $18.25<\bj\leq19.80$ (solid line),
$19.80<\bj\leq20.40$ (dotted line) and $20.40<\bj\leq20.85$ (dashed
line).}
\label{fig_nz}
\end{figure}

\begin{figure*}
\centering
\centerline{\psfig{file=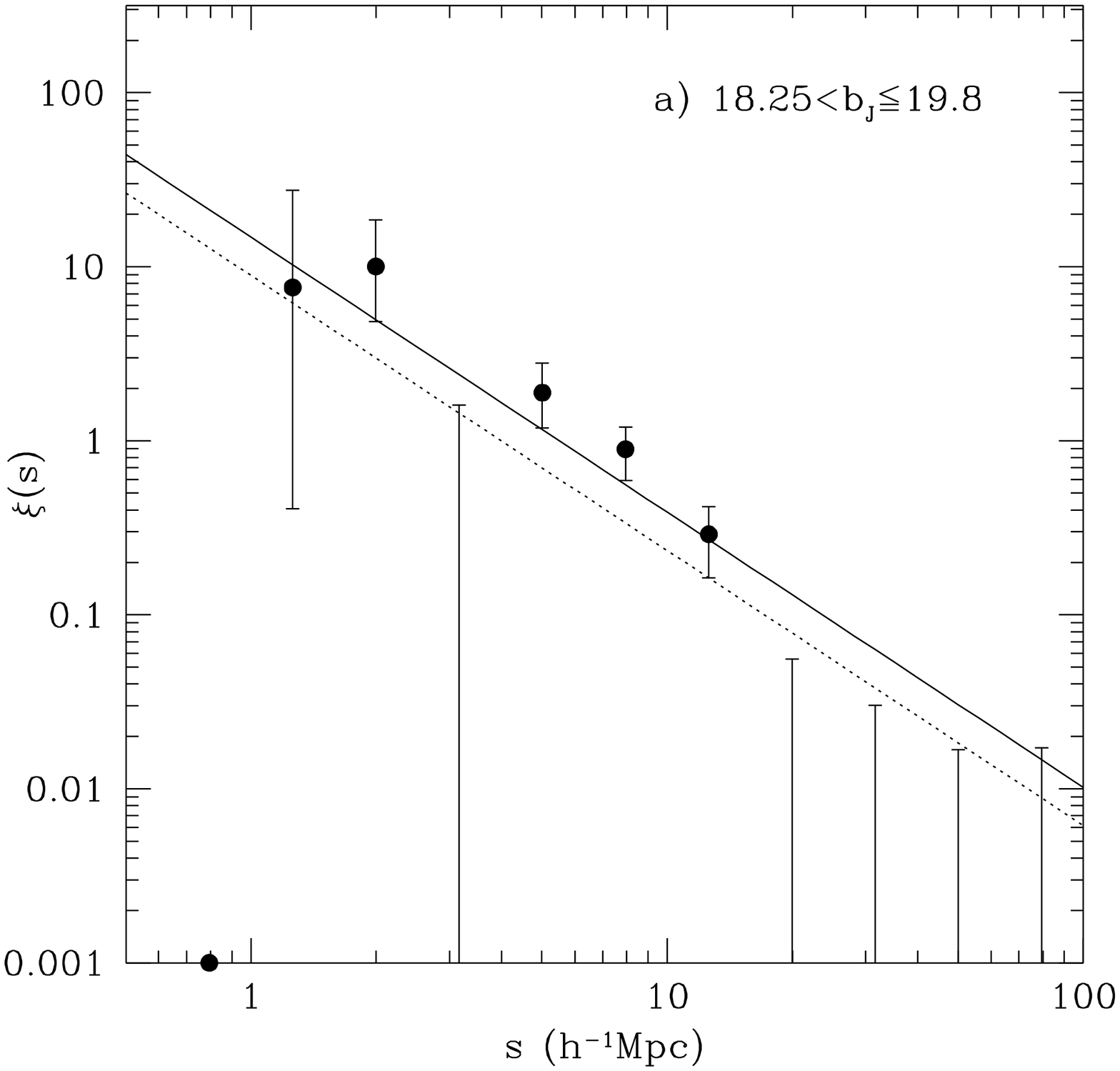,width=8.0cm}}
\centerline{\psfig{file=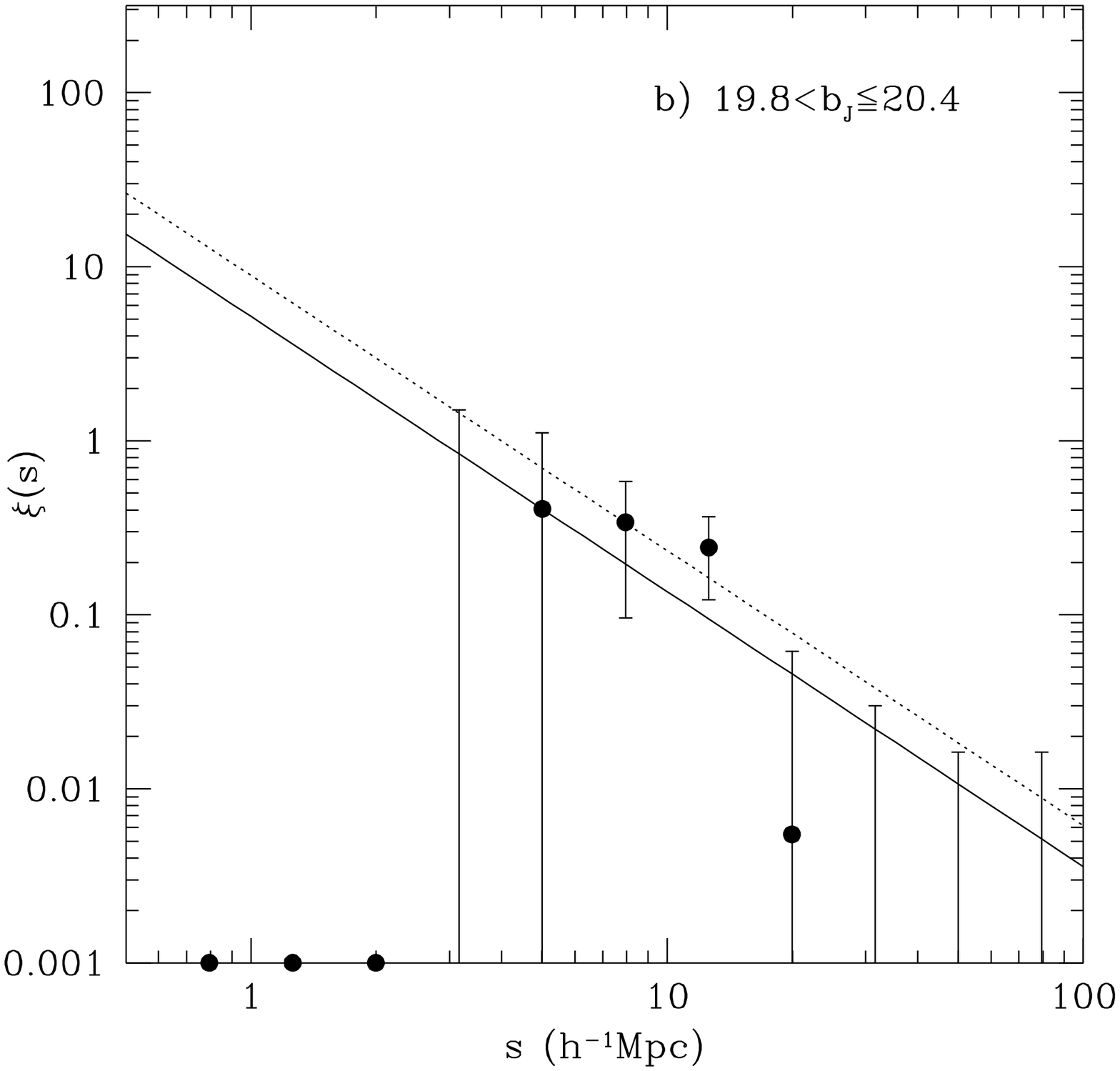,width=8.0cm}\psfig{file=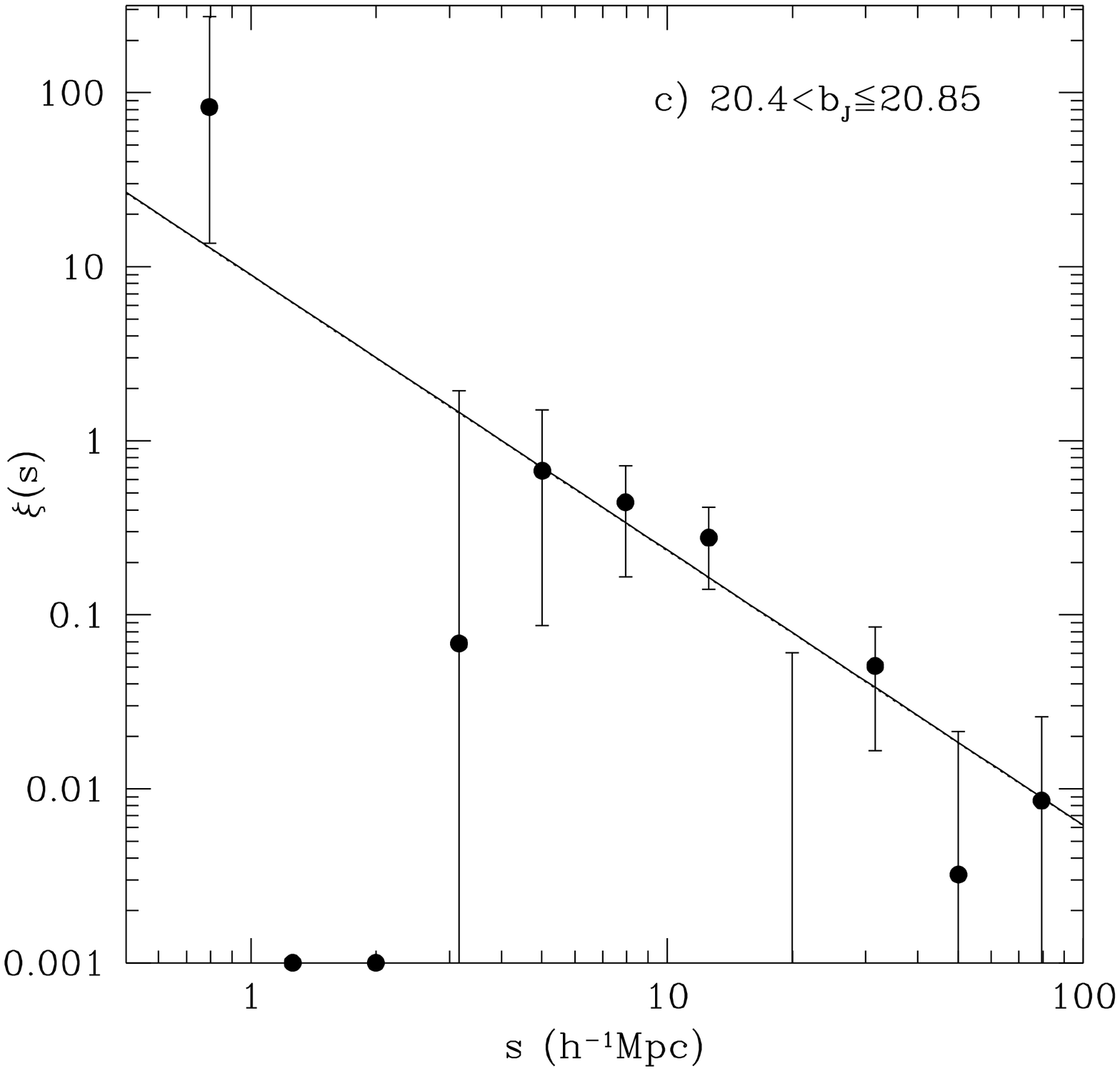,width=8.0cm}}
\caption{The clustering of QSOs as a function of limiting apparent
magnitude in an EdS universe.  a) bright QSOs with
$18.25<\bj\leq19.8$, b) intermediate QSOs with $19.8<\bj\leq20.4$ and
c) faint QSOs with $20.4<\bj\leq20.85$.  The solid line is the best
fit power law in each case (assuming a slope of $-1.58$).  The dashed
line is shown for reference and is the best fit power law for the full
sample from Paper II.  The points at $\xi(s)=0.001$ with no errors
denote bins where no QSO-QSO pairs were found.  These are properly
taken into account in the fitting process.}
\label{fig_xir}
\end{figure*}

The division of QSOs on the basis of their apparent magnitude has
several advantages over using absolute luminosities.  First, apparent
magnitude is a measured, not derived, quantity and is thus not
dependent on the cosmological model used.  It should be easy to produce
flux limited samples from any model of QSO formation, allowing direct
comparison between data and models.  Second, because $M^*$ is at
approximately the same apparent magnitude at every redshift, an
apparent magnitude cut selects the same part of the QSO LF at each
redshift.  

Measuring the clustering of QSOs as a function of luminosity is
challenging.  In particular, at luminosities brighter that $M^*$ the
number density of QSOs falls off as a steep power law.  Splitting a
sample into $N$ redshift bins of equal numbers of QSOs increases the
error on pair counts by $\sim\sqrt{N}$.  Dividing the sample into $N$
luminosity bins increases errors by $\sim N\sqrt{N}$, as the surface
density of sources is also reduced.  This is a further reason why we
limit our analysis in this paper to clustering as a function of
apparent magnitude, as to carry out a meaningful sub-division on the
basis of absolute magnitude requires a larger number of bins (due to
the redshift-luminosity degeneracy inherent in the data set).  A
method to reduce the error in the measured clustering as a function of
absolute luminosity is to cross-correlate QSOs of a particular
luminosity with all other QSOs. This method will be investigated by
Loaring et al. (in preparation).

\subsection{Correlation function estimates}

Our correlation function estimation is carried out as described in
Paper II.  For completeness we outline the procedure here.  The QSO
correlation function, $\xi\qso(s)$, where $s$ is the redshift-space
separation of two QSOs, is calculated using two representative
cosmologies, the EdS and $\Lambda$ models.  The minimum variance
estimator of Landy \& Szalay (1993) is used to derive $\xi\qso(s)$.
In this Paper we divide the 2QZ 10k sample into 3 apparent magnitude
intervals with approximately equal numbers of QSOs.  The magnitude
ranges, mean redshifts and other parameters for these bins are listed
in Table \ref{table_fits}.  We note that the mean redshifts of the
different magnitude slices are very similar, with $\bar{z}=1.398$,
1.509 and 1.552 for the bright, middle and faint slices respectively.
These small differences cannot introduce a significant variation in
clustering, particularly as we find that QSO clustering does not
evolve significantly with redshift (Paper II).  The redshift
distributions are shown in Fig. \ref{fig_nz}.

We correct for the current incomplete observational coverage of the
survey by using a random catalogue which exactly traces the
distribution of observed QSOs on the sky, as in Paper II.  The
redshift distribution of these random points is taken from a spline
fit to the QSO $n(z)$ distribution (each magnitude slice is fit
separately).  We also normalize the number of  randoms within each
UKST field in the survey to remove any possible systematic errors due
to zero-point calibration errors.

We calculate the errors on $\xi\qso$ using the Poisson estimate of
\begin{equation}
\Delta\xi\qso(s)=\frac{1+\xi(s)}{\sqrt{QQ(s)}}.
\label{xierr}
\end{equation}
At small scales, $\lsim50\Mpc$, this estimate is accurate because each
QSO pair is independent (i.e. the QSOs are not generally part of
another pair at scales smaller than this).  On scales larger than
$\sim50\Mpc$ the QSO pairs become more correlated and we use the
approximation that $\Delta\xi\qso(s)=[1+\xi\qso(s)]/\sqrt{N\qso}$,
where $N\qso$ is the total number of QSOs used in the analysis
\cite{sb94,cs96}, for bins in which $QQ(s)>N\qso$.  Note that in this
paper we are concerned with analysis on small scales ($\leq25\Mpc$),
where  the Poisson error estimates are applicable.  The $\sqrt{N\qso}$
errors are only used for displaying our correlation functions.  On
very small scales the number of QSO-QSO pairs can be $\lsim10$.  In
this case simple {\it root-n} errors (Eq. \ref{xierr}) do not give the
correct upper and lower confidence limits for a Poisson distribution.
We use the formulae of Gehrels (1986) to estimate the Poisson
confidence intervals for one-sided 84\% upper and lower bounds
(corresponding to $1\sigma$ for Gaussian statistics).  These errors
are applied to our data for $QQ(s)<20$.  By this point root-n errors
adequately describe the Poisson distribution.

\begin{figure*}
\centering
\centerline{\psfig{file=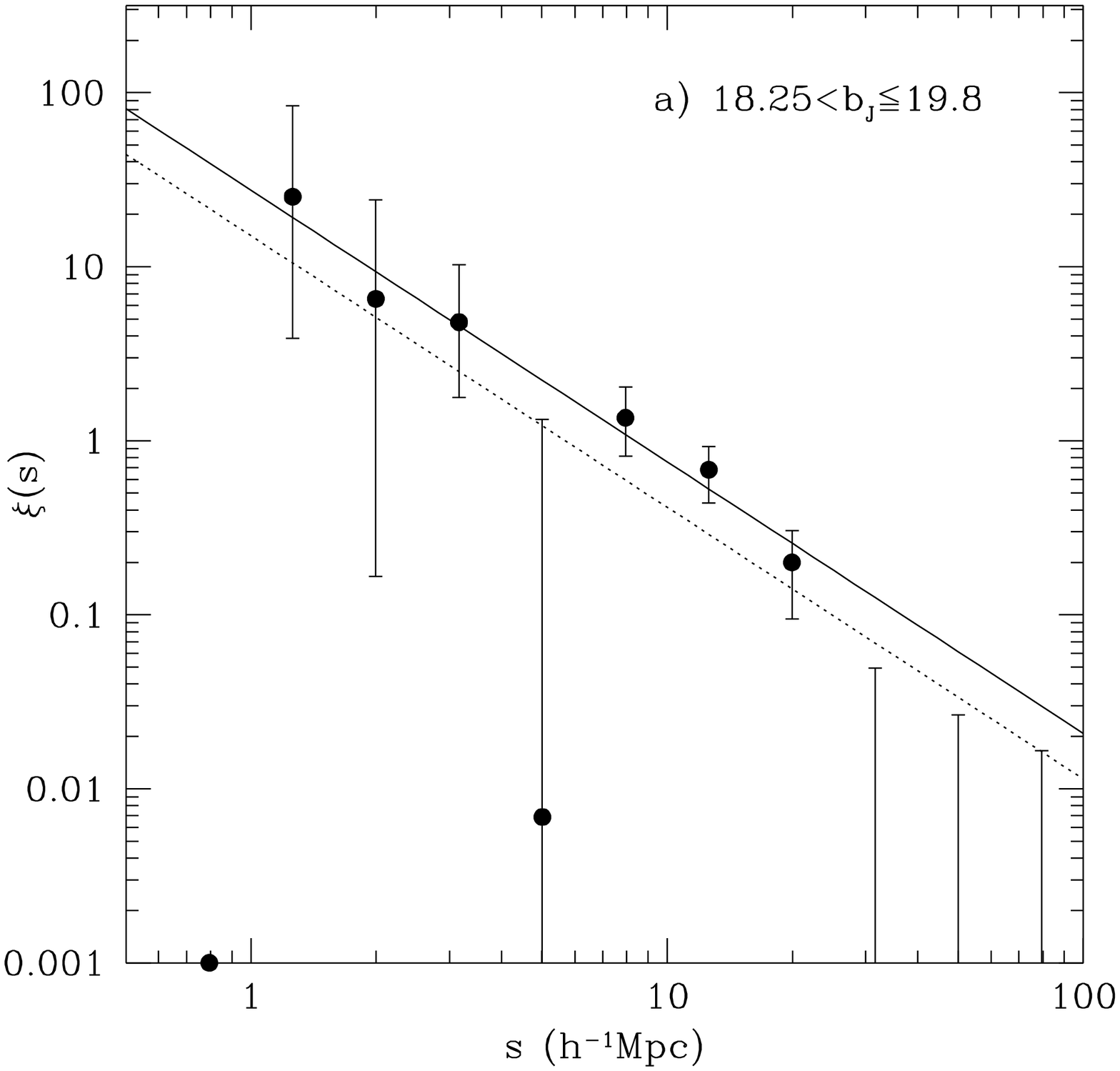,width=8.0cm}}
\centerline{\psfig{file=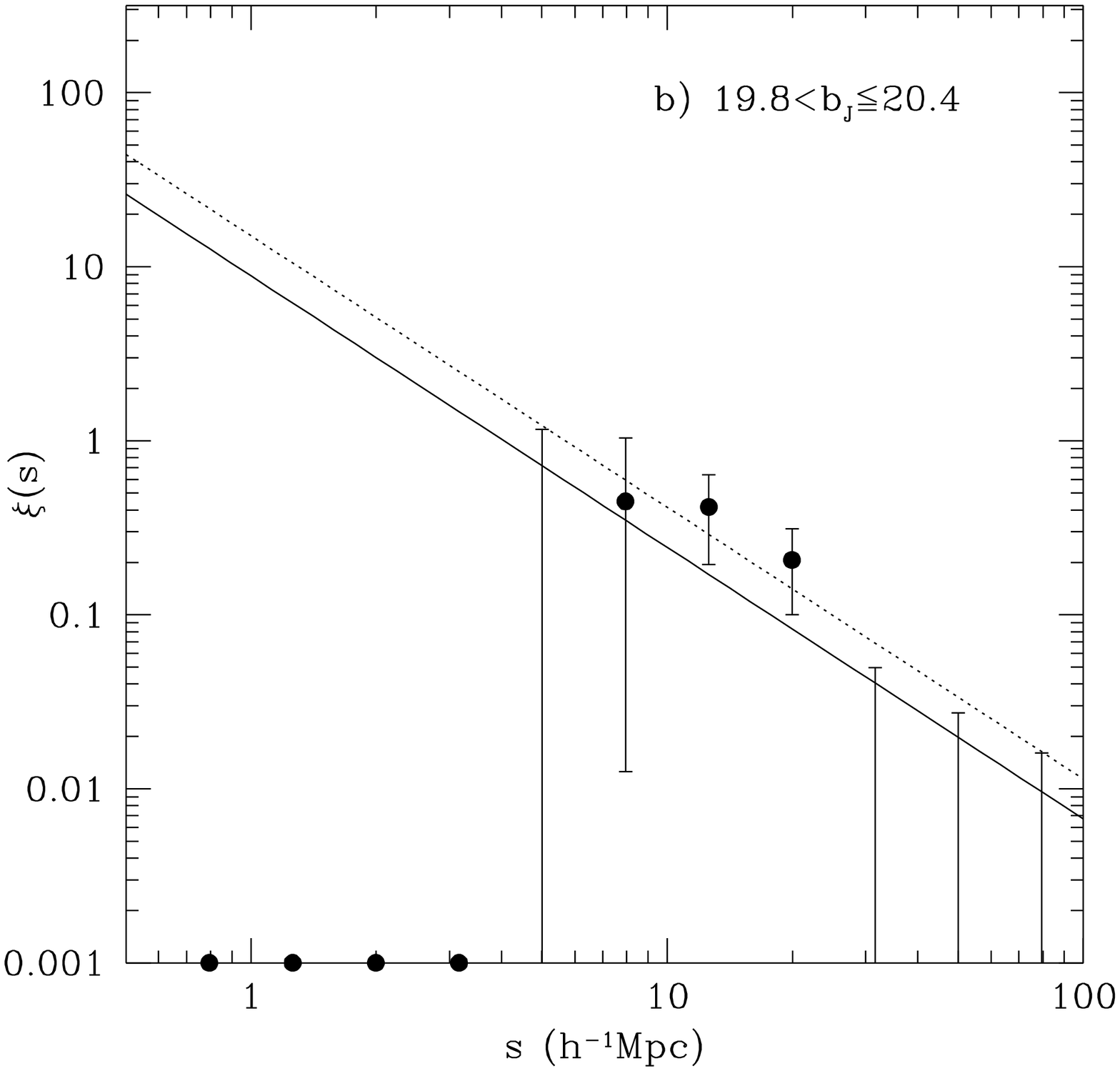,width=8.0cm}\psfig{file=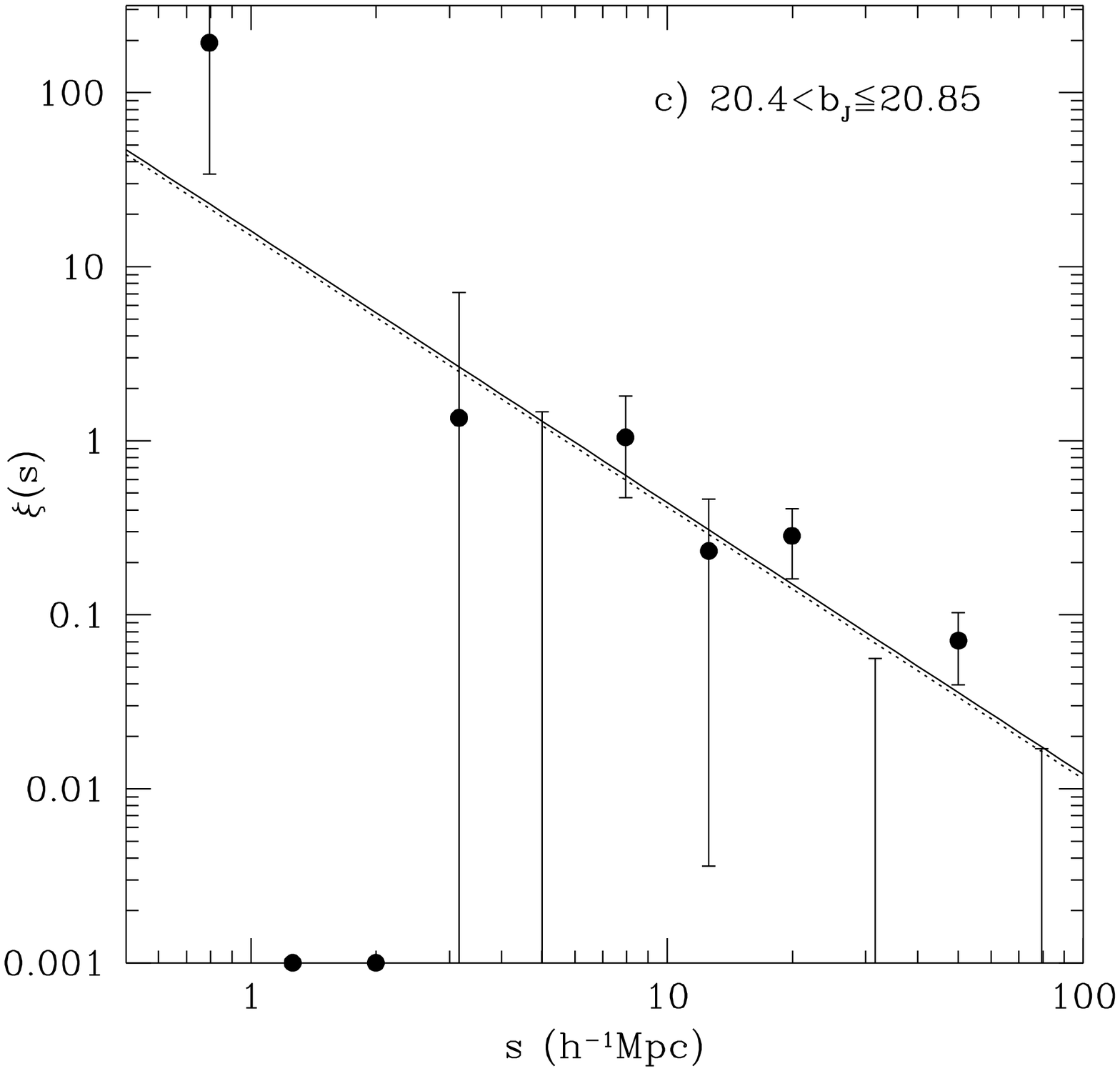,width=8.0cm}}
\caption{The clustering of QSOs as a function of limiting apparent
magnitude in a $\Lambda$ universe.  a) bright QSOs with $18.25<\bj\leq19.8$, b) intermediate
QSOs with $19.8<\bj\leq20.4$ and c) faint QSOs with
$20.4<\bj\leq20.85$.  The solid line is the best fit power law in each
case (assuming a slope of $-1.56$).  The dashed line is shown for
reference and is the best fit power law for the full sample from Paper
II.}
\label{fig_xirlam}
\end{figure*}

\section{QSO clustering as a function of luminosity}\label{sec_results} 

In Fig. \ref{fig_xir} we show the clustering of 2QZ QSOs for three
different apparent magnitude intervals (see Table \ref{table_fits}) in
an EdS universe.  The best fit to the total sample found in Paper II
is shown as the dotted line for reference.  On scales $<20\Mpc$ it
appears that the brighter QSOs (Fig. \ref{fig_xir}a) show marginally
higher clustering that fainter QSOs (Figs. \ref{fig_xir}b and c).  In
order to determine the significance of this difference we make two
measurements.  We first determine the integrated correlation function,
$\xibar\qso$, within a radius  $s\max$, which is defined as
\begin{equation}
\xibar\qso(s\max)=\frac{3}{s\max^3}\int_0^{s\max} \xi\qso(x)x^2{\rm d}x.
\label{xibar_eq}
\end{equation}
We use $s\max=20\Mpc$ as in Paper II.  The results of this are shown
in Table \ref{table_fits}.  While $\xibar\qso(20)$ is greatest for the
bright sample, it is not significantly larger than the values found
for fainter QSOs.  

While $\xibar\qso$ is a robust statistic, it does not take into
account the the shape of $\xi\qso$.  It is possible that a stronger
constraint may be obtainable by fitting a functional form to the data.
We fit a standard power law  of the form
$\xi\qso(s)=(s/s_0)^{-\gamma}$ to our results.  We make the assumption
that the slope of the power law does not vary with luminosity and fix
the slope of  the power law, to be the best fit from the
full data set, that is, $\gamma=1.58$ (EdS) or $\gamma=1.56$
($\Lambda$).  We use a maximum likelihood estimator based  on the
Poisson probability distribution function, so that
\begin{equation}
L=\prod_{i=1}^{N}\frac{e^{-\mu}\mu^{\nu}}{\nu!}
\end{equation}
is the likelihood, where $\nu$ is the observed number of QSO-QSO
pairs, $\mu$ is the expectation value for a given model and $N$ is the
number of bins fitted.  We fit the data with bins $\Delta\log(r)=0.1$,
although we note that varying the bin size by a factor of two makes no
noticeable difference to the resultant fit.  In practice we minimize
the function $S=-2{\rm ln}(L)$, and determine the errors from the
distribution of $\Delta S$, where $\Delta S$ is assumed to be
distributed as $\chi^2$.  This procedure does not give us an absolute
measurement of the goodness-of-fit for a particular model.  We
therefore also derive a value of $\chi^2$ for each model fit in order
to confirm that it is a reasonable description of the data.  We carry
out the fit on scales $0.7<s\leq20\Mpc$, the scales chosen being the
smallest scale containing a QSO-QSO pair and the scale at which
$\xi\qso$ appears to break from a power law.

The best fit models are show in Fig. \ref{fig_xir} (solid lines).  As
suggested by the $\xibar\qso$ measurements, the brightest sample does
show stronger clustering than the other samples, however the
difference is only marginally significant.  It should also be noted
that the lowest clustering amplitude is found in the intermediate
sample, although the fit to this data set is consistent with that of
the faintest sample.

In Fig. \ref{fig_xirlam} we plot the clustering of QSOs in a $\Lambda$
universe.  This shows the same feature as the EdS case, with the
brightest QSOs being the most strongly clustered.  The measured
$\xibar\qso(20)$ values (Table \ref{table_fits}) also show this.  The
significance of the difference between the faintest and brightest
sub-samples is now larger, but still only at the $\sim1.7\sigma$
level.  We carry out the power law fit (this time between
$0.8<s\leq25\Mpc$) as above (solid lines), again finding only a
marginal difference between different magnitude intervals.

\begin{figure}
\centering
\centerline{\psfig{file=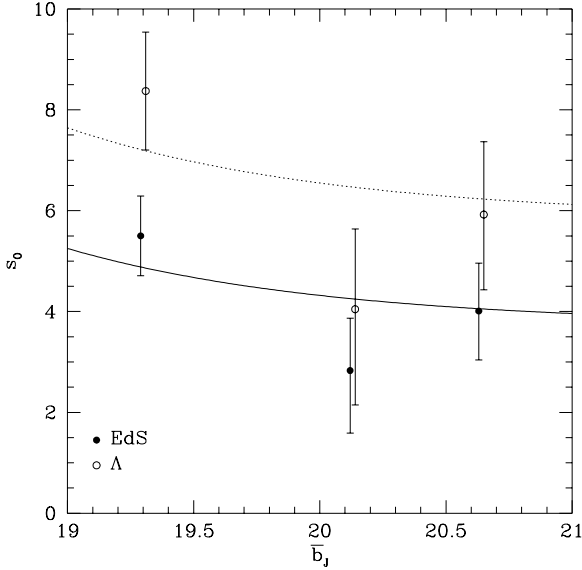,width=8.0cm}}
\caption{The QSO clustering scale length, $s_0$ as a function of mean
apparent magnitude in an EdS universe (filled circles) and a $\Lambda$
universe (open circles).  The best fit galaxy luminosity dependence
models (see Section \ref{sec_discuss}) for EdS (solid line) and
$\Lambda$ (dotted line) cosmologies are also shown.}
\label{fig_r0}
\end{figure}

In Fig. \ref{fig_r0} we plot the fitted values of $s_0$ as a function
of mean apparent magnitude.  In order to show a convincing trend as a
function of magnitude we would like to span a much larger range in
magnitude than is possible with the current sample.

\section{Discussion}\label{sec_discuss}

Fitting the full data set over the same range of scales as above
we find that $s_0=4.29^{+0.30}_{-0.29}\Mpc$ (EdS) and
$s_0=6.351^{+0.45}_{-0.44}\Mpc$ ($\Lambda$).  Thus the brightest third
of QSOs has a clustering scale length which is a factor of
$1.28\pm0.20$ (EdS) or $1.32\pm0.20$ ($\Lambda$) larger than that of
the full sample.  As $b/b^*=(s_0/s_0^*)^{\gamma/2}$ this then
implies that the ratio of the biases is $1.22\pm0.15$ (EdS) or
$1.24\pm0.15$ ($\Lambda$).  We therefore find only weak evidence that bright
QSOs cluster more strongly.

We should also consider comparing the above results to other
measurements of luminosity dependent clustering.  The most accurate
measurement of this to date is by  Norberg et al. (2001) using data
from the 2dF Galaxy Redshift Survey.  Norberg et al. find that the
clustering of $z<0.3$ galaxies is a weak function of luminosity
fainter than $L\gal^*$, but a much stronger function of luminosity for
galaxies brighter than $L\gal^*$.  They find that the relation
$b\gal/b\gal^*=0.85+0.15L\gal/L\gal^*$ well describes the luminosity
dependence of galaxy bias, $b\gal$, relative to that found for
$L\gal^*$ galaxies.  If there was a simple one-to-one relation between
galaxy luminosity and QSO luminosity, we would be able to make a
straightforward comparison of QSO clustering to this relation.
Unfortunately, we know that there is only a weak correlation (with
large dispersion) between these two quantities \cite{sbl00}.  We can,
however, at least check for consistency between the galaxy and QSO
clustering luminosity dependence.  Under the assumption that galaxy
luminosity is more closely related to halo mass than QSO luminosity
is, we would expect the change in QSO clustering with luminosity to be
no more rapid than the change in galaxy clustering with luminosity.
Converting the above relation for galaxy clustering to a form directly
applicable in our case we find
$(s_0/s_0^*)^{\gamma/2}=0.85+0.15\times10^{-0.4(m-m^*)}$, where $m^*$
is the apparent magnitude corresponding to $M^*$.  The value of $m^*$
varies in the range 19.3 to 19.6 depending on the particular best fit
LF model used.  We assume throughout that $b$ is not a function of
scale and allow $s_0^*$ to be a single free parameter, while fixing
$m^*$ [$m^*=19.6$ (EdS) or 19.3 ($\Lambda$)].  The best fit values are
$s_0^*=4.78\pm0.57\Mpc$ (EdS) or  $s_0^*=7.20\pm0.86\Mpc$ ($\Lambda$)
and lines corresponding to these are shown in Fig. \ref{fig_r0}.  We
find that the above empirical model for galaxies is also consistent
with the clustering of QSOs in the 2QZ.  However, this is
un-surprising given the large errors on the QSO clustering
measurements.
  
A number of authors have attempted to model the clustering of QSOs,
based on the Press-Schechter (1974) formalism for describing the
evolution of dark matter haloes \cite{mw01,hh01}.  These make the
assumption that the luminosity of QSOs is correlated with mass of dark
matter halo in which they reside.  Although there is  reasonable
evidence that the {\it mass} of compact objects (presumably black
holes) is correlated with the mass of the spheroidal component of the
host galaxy (e.g. Magorrian et al. 1998), a correlation between AGN
{\it luminosity} and host luminosity and/or mass is less evident.  The
time-scale of QSO activity is the main parameter that controls
amplitude of clustering in these models, with time-scales of order
$10^6$ years being most consistent with the results found in Paper II
\cite{kh01}  The main arguments are based on the number-density of
galaxies that have gone through a QSO phase.  Similar number-density
arguments suggest that, as brighter QSOs are rarer, they should cluster
more strongly.  Our current analysis does not have sufficient
signal-to-noise to clearly demonstrate this, and a more detailed
investigation will have to  await brighter QSO surveys such as the
Sloan Digital Sky Survey \cite{sdssqso} and the bright extension to
the 2QZ being carried out with FLAIR and the 6-degree Field system on
the UK Schmidt Telescope.  Further investigation is also possible
using other techniques, such as the {\it cross}-correlation of QSOs of
different luminosities.  This will be investigated in the 2QZ by
Loaring et al. (in preparation).

\section*{acknowledgments}

We warmly thank all the present and former staff of the
Anglo-Australian Observatory for their work in building and operating
the 2dF facility.  The 2QZ is based on observations made with the
Anglo-Australian Telescope and the UK Schmidt Telescope.  NSL is
supported by a PPARC Studentship.


\begin{thebibliography}{}

\bibitem[Andreani \& Cristiani 1992]{ac92}
Andreani P.,  Cristiani S.,  1992, ApJ, 398, L13

\bibitem[Boyle et al. 2000]{2qzpaper1}
Boyle B.~J.,  Shanks T.,  Croom S.~M.,  Smith R.~J.,  Miller L.,
Loaring N., Heymans C.,  2000, \mnras, 317, 1014

\bibitem[Cristiani \& Vio 1990]{cv90} 
Cristiani S.,  Vio R.,  1990, \aap, 227, 385

\bibitem[Croom \& Shanks 1996]{cs96}
Croom S.~M.,  Shanks T.,  1996, MNRAS, 281, 893

\bibitem[Croom et al. 2001a]{2qzpaper2}
Croom S.~M.,  Shanks T.,  Boyle B.~J.,  Smith R.~J.,  Miller L.,  Loaring N.,
   Hoyle F.,  2001a, \mnras, 325, 483

\bibitem[Croom et al. 2001b]{2qzpaper5}
Croom S.~M.,  Smith R.~J.,  Boyle B.~J.,  Shanks T.,  Loaring N.~S.,  Miller
  L.,    Lewis I.~J.,  2001b, \mnras, 322, L29

\bibitem[Fan et al. 2001]{sdsslf01}
{Fan} X. et al.,  2001, \aj, 121, 54

\bibitem[Gehrels 1986]{g86}
Gehrels N.,  1986, \apj, 303, 336

\bibitem[Giavalisco \& Dickinson 2001]{gd01}
Giavalisco M.,  Dickinson M.,  2001, ApJ, 550, 177

\bibitem[Haiman \& Hui 2001]{hh01}
Haiman Z., Hui L., 2001, ApJ, 547, 27

\bibitem[Hewett et~al. 1995]{lbqs6}
Hewett P.~C.,  Foltz C.~B.,    Chaffee F.~H.,  1995, AJ, 109, 1499

\bibitem[Iovino \& Shaver 1988]{is88}
{Iovino} A.,  {Shaver} P.~A.,  1988, \apjl, 330, L13

\bibitem[Kauffmann \& Haehnelt 2001]{kh01} Kauffmann G., Haehnelt M.,
2001, \mnras\ submitted (astro-ph/0108275)

\bibitem[La Franca et~al. 1998]{lac98}
{La Franca} F.,  {Andreani} P.,    {Cristiani} S.,  1998, \apj, 497, 529


\bibitem[Landy \& Szalay 1993]{ls93}
{Landy} S.~D.,  {Szalay} A.~S.,  1993, \apj, 412, 64

\bibitem[Loveday et~al. 1995]{lmep95}
Loveday J.,  Maddox S.~J.,  Efstathiou G.,    Peterson B.~A.,  1995, ApJ, 442,
  457

\bibitem[Magorrian et al. 1998]{mag98}
Magorrian J. et al.,  1998, \aj, 115, 2285

\bibitem[Martini \& Weinberg 2001]{mw01}
Martini P., Weinberg D.~H., 2001, ApJ, 547, 12

\bibitem[Norberg et al., 2001]{nor01}
Norberg P. et al., 2001, \mnras\ submitted (astro-ph/0105500)

\bibitem[Press \& Schechter 1974]{ps74}
Press W.~H., Schechter P., 1974, ApJ, 187, 425

\bibitem[Schade et~al. 2000]{sbl00}
{Schade} D.~J.,  {Boyle} B.~J.,    {Letawsky} M.,  2000, \mnras, 315, 498

\bibitem[Schneider et~al. 2002]{sdssqso}
Schneider D.~P. et al., 2002, AJ, 123, 567

\bibitem[Shanks \& Boyle 1994]{sb94}
Shanks T.,  Boyle B.~J.,  1994, MNRAS, 271, 753

\end{thebibliography}
\end{document}